%% file: 0_main.tex
\documentclass[manuscript]{acmart}

\usepackage{soul}
\usepackage{tfrupee}
\usepackage[normalem]{ulem}
\usepackage{natbib}

\usepackage{xcolor}
\definecolor{control}{RGB}{0, 76, 109}
\definecolor{intervention}{RGB}{105, 150, 179}
\definecolor{white}{RGB}{255, 255, 255}
\definecolor{black}{RGB}{0, 0, 0}
\definecolor{median}{RGB}{237, 157, 66}

\begin{document}

%Part of this issue is due to the Widely available MT tools are not designed specifically for use in healthcare, and evaluations of these systems on medical language suggests a potential for grave harm from translation errors \cite{Vieira21}. 
\title{Beyond General Purpose Machine Translation: The Need for Context-specific Empirical Research to Design for Appropriate User Trust}

%\title{The need for Empirical Research under specific context to build Appropriate Trust in Machine Translation Systems}

%\
% Attending to the context of use to inform the design of trustworthy machine translation
% The need for context-specific empirical research to inform the design of trustworthy machine translation
% The need for Empirical Research in specific contexts to inform the design of Machine Translation Systems that foster Appropriate user Trust.
% The context-specific nature of trust in MT systems
% Designing for appropriate user trust in MT requires empirical research into specific contexts of use
% Beyond general purpose Machine Translation (MT): the need for context-specific empirical research and system design to calibrate appropriate user trust in MT

\author{Wesley Hanwen Deng}
\affiliation{%
  \institution{Carnegie Mellon University}
  \city{Pittsburgh}
  \state{Pennsylvania}
  \country{USA}}
\email{hanwend@cs.cmu.edu}

\author{Nikita Mehandru}
\affiliation{%
  \institution{University of California, Berkeley}
  \city{Berkeley}
  \state{California}
  \country{USA}}
\email{nmehandru@ischool.berkeley.edu}

\author{Samantha Robertson}
\affiliation{%
  \institution{University of California, Berkeley}
  \city{Berkeley}
  \state{California}
  \country{USA}}
\email{samantha_robertson@berkeley.edu}

\author{Niloufar Salehi}
\affiliation{%
  \institution{University of California, Berkeley}
  \city{Berkeley}
  \state{California}
  \country{USA}}
\email{nsalehi@berkeley.edu}

\renewcommand{\shortauthors}{TBA}

\begin{abstract}
\input{sections/00_abstract.tex}
\end{abstract}

%\begin{CCSXML}
%<ccs2012>

%<concept>
%<concept_id>10003120.10003130.10011762</concept_id>
%<concept_desc>Human-centered computing~Empirical studies in collaborative and social %computing</concept_desc>
%<concept_significance>500</concept_significance>
%</concept>

%<concept>
%<concept_id>10003120.10003121</concept_id>
%<concept_desc>Human-centered computing~Human computer interaction (HCI)</concept_desc>
%<concept_significance>500</concept_significance>
%</concept>

%<concept>
%<concept_id>10003120.10003121.10003122.10003334</concept_id>
%<concept_desc>Human-centered computing~User studies</concept_desc>
%<concept_significance>100</concept_significance>
%</concept>

%</ccs2012>
%\end{CCSXML}

%\ccsdesc[500]{Human-centered computing~Empirical studies in collaborative and social computing}
%\ccsdesc[500]{Human-centered computing~Empirical studies in HCI}
%\ccsdesc[500]{Human-centered computing~Human computer interaction (HCI)}

%\keywords{TBA}

\maketitle

\section{Introduction}
\input{sections/01_introduction.tex}

\section{Five challenges in building trustworthy machine translation} \label{2}
\input{sections/02_challenge.tex}
\section{Case Study: Using Machine Translation in Medical Settings}
\input{sections/03_case.tex}
\section{Towards trustworthy Machine Translation systems} \label{4}
\input{sections/04_future.tex}

\clearpage
\bibliographystyle{ACM-Reference-Format}
\bibliography{citation}

\end{document}

%% file: sections/00_abstract.tex
Machine Translation (MT) has the potential to help people overcome language barriers and is widely used in high-stakes scenarios, such as in hospitals. However, in order to use MT reliably and safely, users need to understand when to trust MT outputs and how to assess the quality of often imperfect translation results. %More research is needed to study and design for building and calibrating trust in MT systems. 
In this paper, we discuss research directions to support users to calibrate trust in MT systems. We share findings from an empirical study in which we conducted semi-structured interviews with 20 clinicians to understand how they communicate with patients across language barriers, and if and how they use MT systems. Based on our findings, we advocate for empirical research on how MT systems are used in practice as an important first step to address the challenges in building appropriate trust between users and MT tools. 

%% file: sections/01_introduction.tex
Machine Translation (MT) is widely available as stand-alone translation tools \cite{Wu2016GooglesNM, googletranslate} and embedded in larger user-facing systems \cite{facebooktranslate, twittertranslate}. Recent advances in neural machine translation have accelerated the progress of MT design and development \cite{Wu2016GooglesNM, Bahdanau2015NeuralMT}. Despite the rapid technological improvement, however, MT models can still produce translations that are imperfect or even wrong, causing frustrations or harm to individuals when people over-rely on the outputs \cite{Yamashita2006EffectsOM, Yamashita2009DifficultiesIE}. For instance, in 2017, the MT system embedded in Facebook translated a post from a Palestinian man saying ``good morning'' in Arabic to ``attack them,'' and the poster was arrested by Israeli police \cite{berger2017attack}. As MT is widely used in high-stakes contexts such as healthcare \cite{liu2019comparison}, immigration \cite{stecklow2018facebook}, and law \cite{Vieira21}, it is important to develop ways to support end-users to build appropriate trust in MT systems \cite{robertson_2021_directions}. Calibrating trust and reliance could enable more effective human-MT collaboration. At the same time, MT systems should encourage users to fully understand the limits of translation technology to avoid harm caused by over-trust. %However, there exists limited research synthesizing the challenges involved in building appropriate user trust in MT systems.

In this paper, we first discuss five challenges in calibrating trust between users and MT systems. In particular, users lack intuition and expertise to guide their judgments of translation quality, since they usually have no or limited proficiency in one of either the source or the target language. In addition, populations that might benefit from trustworthy MT systems the most could be hard to reach and engage equitably in research. Adding an additional layer of complexity, it is difficult to communicate translation quality to end-users, since  translation is nuanced and context-dependent.

We then present findings from an empirical study in which we conducted semi-structured interviews with 20 clinicians to understand how MT was used in their daily practices and their needs for trust in medical MT systems. We observed that MT served as a low-cost and efficient way of facilitating conversations between clinicians and patients in under-resourced hospitals, but was usually viewed as a last-resort option due to the lack of trust, especially in high-stakes medical settings where patients’ consent was required. We also identified techniques clinicians had developed to navigate and calibrate trust in communications with patients that were mediated by human interpreters, which offer some paths forward on designing future trustworthy MT tools. In addition, while current commercial MT systems are mostly general-purpose, clinicians were vocal about having MT systems that are specifically designed to prioritize accurate translation in medical settings and evaluated by a hospital board or medical society.

To conclude, we stress the need for conducting empirical studies to understand how people navigate language barriers and whether and how MT systems are being used in specific domains and contexts. This understanding is an important but currently neglected first step to address the challenges in building appropriate trust between users and MT tools.

%\sr{Could mention that we actually know very little about how MT is used because companies aren't sharing that info, but that there is evidence that it is used in these high-stakes domains.}

% Ken told me it's fine to be less rigorous on the definition of human-AI teaming :-).
%When being used in the real-world, MT systems often form a "team" with the users, situating in the specific environment and context \cite{suchman2007human}. For example, people commonly use MT systems to help interpret articles written in a language they don't read based on MT's output. Similarly, text-based online communications between two people who don't understand each other's languages are often mediated by MT. 

%% file: sections/02_challenge.tex
Prior research has surfaced several reasons why developing trust in AI systems, and large language models in particular, is hard. For example, model evaluations based on metrics like the BLEU score \cite{Papineni2002BleuAM} is disconnected from real-world application performance that is related to how users perceive the outcome qualities \cite{Post2018ACF}. This gap between evaluation and performance is exacerbated by design asymmetries between developers and users in the current AI product lifecycle. In particular, large language models are usually constructed by practitioners with access to substantial computational resources \cite{bender2021dangers}, users whose lives are impacted by these models are being left out in the design and development \cite{sap2019social}. Current MT systems based on neural machine translation share this challenge when it comes to calibrating trust between MT systems and users. 

In this section, we further detail five distinct challenges that exist in building trustworthy MT systems, based on prior literature from both the MT and HCI communities and our own experience conducting empirical studies on MT systems \cite{robertson_2021_directions, robertson_2022_understanding, mehandru_2022_reliable}.

\subsection{Users lack intuition and expertise to guide their judgments of translation quality} \label{2.1}

A central challenge in designing trustworthy machine translation systems for everyday users is that these users are usually using MT because they are not fluent in one of either the source or the target language. This means that they have \textbf{limited ability to assess the quality of the translations that an MT system has produced and determine whether to rely on them or not.} Prior research studying trust in ML systems and developing interventions to appropriately calibrate trust has largely focused on systems where the user either has expertise in the decision domain and is being assisted by ML, or where some general knowledge or intuition applies \cite{suresh2021beyond}. MT poses unique challenges because users often cannot comprehend the output of the system, let alone how it was produced or whether it is correct. To circumvent this challenge, users and researchers have relied heavily on back-translation (translating an output back to the source language) \cite{shigenobu2007backtranslation,miyabe2010detecting,yamashita2009difficulties,miyabe2009backtranslation}. While this offers some insight into the translation in a language the user can understand, it remains unclear how reliable this strategy is in practice, or in what specific cases it helps or fails to support users in calibrating their trust in MT \cite{miyabe2010detecting, zouhar-etal-2021-backtranslation}. For example, Zouhar et al. found that showing users the backtranslation increased their trust in an MT model, but did not help them to identify or improve poor quality translations \cite{zouhar-etal-2021-backtranslation}. This is consistent with evidence in MT and in other ML domains that showing additional interpretability or explainability information can increase users' trust in a model simply because the additional information is there, even if that information indicates that the model may be incorrect \cite{miyabe2010detecting, kaur2020interpreting}.

Researchers have also designed and evaluated interfaces that provide additional information other than the backtranslation, such as two different translations \cite{gao2015twoOutputs,xu2014improving}, highlights of key words \cite{gao2013highlighting,lim2018sensetrans}, emotional and cultural context \cite{lim2018sensetrans}, and numerical indications of translation quality \cite{miyabe2011can}. Lab studies suggest that additional information can improve message recipients' perceived understanding without increasing the mental workload \cite{gao2013highlighting,lim2018sensetrans,gao2015twoOutputs,xu2014improving}, but it is not clear from these studies whether users' \textit{perceived} understanding aligns with their actual understanding of the intended meaning \cite{robertson_2022_understanding}. In other words, we do not know whether these interventions help users appropriately calibrate their trust in MT, or whether it promotes trust indeterminately even when translations are incorrect. 

\subsection{Translation quality is nuanced and context-dependent} \label{2.2}

To design systems that help a user decide whether to trust a translation or not, researchers need an operational definition of what makes a translation trustworthy. However, translation quality is complex, nuanced, and context-dependent \cite{specia-etal-2013-quest}. \textbf{In many cases, there are many acceptable translations of a single utterance, while in others there is no translation that truly captures the meaning of a word or phrase in the source language.} Whether a user should rely on a particular translation depends heavily on the context of use. For instance, the same translation error may be humorous in one context, but rude in another. In some circumstances, accurate translation of specific terminology may be crucial, while in others users may prioritize getting the gist of an utterance or enough meaning to move on with an interaction.  In each case, the design requirements for trustworthy MT may look very different and it remains unclear when, how, and what kind of quality measures might be intelligible and actionable to users in different situations.

\subsection{Even rare and subtle errors can be extremely detrimental to users} \label{2.3}

Counter-intuitively, another challenge for designing for trust in MT is the relatively high performance of consumer MT systems. Neural MT models have improved substantially in recent years, especially for high resource languages. While we would certainly expect the user experience to improve with higher quality translations, this poses several challenges for engendering appropriate trust. First, as users interact with higher quality translations, their trust in the underlying model is likely to improve. Neural MT models are especially good at producing fluent translations, even if they do not adequately convey the meaning of the source text. Research has suggested that interacting with fluent but inadequate translations does not negatively impact trust, possibly leading users to overtrust a model that produces very fluent output \cite{martindale_fluency_2018}. Second, reports of excellent performance may lead researchers and developers to deprioritize the development of more trustworthy MT systems, believing that there is little need to warn users of errors if those errors are rare and insignificant \cite{bender2021dangers}. 

However, \textbf{particularly in high-stakes settings, rare errors can have drastic consequences}. For instance, Khoong et al. used Google Translate to translate emergency department discharge instructions into Spanish and Chinese \cite{Khoong19}. While the authors found that most translations were acceptable quality, a small number contained potentially life-threatening mistakes, for example, when ``hold the kidney medicine until you have a chance to speak with your kidney doctor,'' was translated to ``keep taking kidney medicine until you talk to your kidney doctor'' in Chinese. Further, MT performance in high-resource languages has advanced far more quickly than performance in lower resource languages. The need to support reliable use of MT in low-resource languages is made even more urgent by this performance disparity, as users may not be aware of these differences and may not adjust their level of trust in the system when switching from a higher-resource to a lower-resource language pair.

% \dhw{I like this point a lot! Also wondering if we could make argument like "the fluency is better which would hide the errors." Not sure if this is the best place to add this.}
% \dhw{Gap between NLP research community norms/priority and the reality performance?}

\subsection{It is challenging to conduct ecologically valid empirical studies of MT} \label{2.4}
Recent studies on trustworthy AI emphasize the importance to design and conduct sound empirical research on how AI systems are being used (and misused) in real-world scenarios in order to capture a well-rounded and comprehensive understanding of the trust construct \cite{gille2020we}. However, \textbf{people who might benefit the most from trustworthy MT are also those who tend to be hard to reach or collaborate with}. For example, when Liebling et al. tried to understand how migrants in India used MT, female participants were reluctant to be interviewed without their husbands present or rejected the interview without permission from their husbands \cite{liebling2020unmetneeds}, biasing their participant sample towards male participants. There is also an epistemological burden for conducting research studies with community members, especially to marginalized social groups or people working in high-stakes jobs, as they are often short on time and resources \cite{pierre2021getting, hsu2021empowering}. This challenge was reflected in one of the authors' study designs in which we attempted to conduct an ethnographic study with COVID testing volunteers who use MT to help communicate with local immigrants. We ultimately decided to change the study design realizing that our study would impose burdens on already over-worked volunteers and their hectic working schedule. 

%However, ecologically evaluating machine translation with real-world stakeholders outside the laboratory environment is difficult. To start with, the use cases of machine translation in daily life are ubiquitous \cite{}, transient \cite{}, and highly contextualized \cite{}. A large body of research focusing on evaluating MT quality and trust was conducted through crowdsourcing platforms like Amazon Mechanical Turks for the convenience and scalability of recruitment \cite{}. The validity of using crowd workers to understand socio-technical perspectives of AI systems, however, was critiqued by scholars \cite{}. Behavior study methods like the semi-structured interview that are conducted under laboratory environments serve as a starting point to understand people's interaction with the MT system, but bear the burden of recall bias \cite{} and observer bias \cite{}.

%In addition, prior work studying MT in medical setting mainly recruited clinicians instead of patients for their user study, largely due to ... \dhw{citation} 

\subsection{High dimensional design space} \label{2.5}

Currently, the most popular machine translation tools, like Google Translate or Microsoft Translator, are general purpose tools that serve a wide range of users involved in a wide range of tasks across a wide range of languages. However, as each of the above challenges highlighted, \textbf{designing for \textit{trustworthy} MT depends heavily on who is using an MT system, and for what}. MT performance varies widely across language pairs, and across dialects of a language \cite{das2019dangers}. Some users have proficiency in source and target language, and are using MT for only a word or phrase they do not know how to translate, while others completely lack knowledge of one of the languages. People use MT in a huge diversity of contexts, from low-stakes settings like reading foreign news or getting directions while traveling, to high-stakes settings like healthcare, content moderation, and policing \cite{Khoong19, torbati2019immigration, vieira2020societal, berger2017attack}. Each of these contexts will call for different kinds of support for users to judge when it is ethical and appropriate to rely on MT. Thus, we have begun by narrowing in to understand how clinicians use MT in their practice, whether they are able to do so reliably, and how we might be able to help them calibrate their trust in translation technologies so that they can provide language support more safely.

% The diversity of language and language translation. 

% - Translation quality itself is very nuanced and domain-specific. It follows that the Quality Estimation (QE) for translation is very difficult and hard to give indications of quality

% - There are so many languages, low-resource languages are understudied. Within each language, there are different dialects, formalities. If someone calibrates their trust in MT with one lang pair, may not apply to another

% Real-world context of using MT

% - People usually can’t understand one of the languages while using the MT, adding extra complexity. 

% - Finally, MT is currently being widely used by marginalized, vulnerable community members like immigrants in high-stake scenarios like healthcare \cite{}, employment \cite{}, and social support \cite{}. However, designing and running studies with real-world stakeholders under high-stake scenarios is challenging…. <...>

%% file: sections/03_case.tex
While professional medical interpreters remain the gold standard for facilitating cross-lingual communication in medical practice, finding skilled translators in a timely manner can be difficult and time-consuming. Clinicians have resorted to machine translation tools when short for time and when operating in low-stakes medical settings \cite{moberly2018choose}. Although there are safety and reliable concerns around machine translation (MT) system deployment in medical settings, MT can be a low-cost and efficient way of communicating with patients in  under-resourced hospitals and clinics where there are limited or no medical interpreters. To minimize risk in this setting, there is a great need to build more trustworthy MT system to facilitate the conversation between clinicians and patients when they don't speak the same language.

To this end, we conducted an interview study with 20 clinicians, including physicians, surgeons, nurses, and midwives, in the U.S., as an initial step to understand how clinicians are currently interacting with machine translation systems. Our study included clinicians from across seven medical specialities: cardiology, orthopedic surgery, nephrology, family medicine, obstetrics and gynecology, trauma surgery, and emergency medicine.  We found that medical literacy rates among patients, cultural barriers, and time and resource constraints greatly impact patient-clinician communication. Building trustworthy systems that ultimately account for reliability and transparency can improve the quality of care of patients in low-resource settings, and can complement existing language translation services in high-resource settings \cite{khoong2022research}. Our findings have important implications for how we might address the challenges above in the design of trustworthy machine translation systems for clinician use.

\subsection{Cross-lingual communication Medical Settings}

Generally, clinicians in our sample turned to professional medical interpreters when they needed to communicate with a patient with whom they did not share a language. However, challenges like time constraints and limited interpreter availability led some to occasionally resort to machine translation. In this section, we outline how and when clinicians used MT in their practice. In the next section, we explore how they tried to evaluate and calibrate their trust in both human interpreters and MT and identify unmet needs and opportunities for designing trustworthy translation support for medical settings.

Some clinicians only used machine translation tools when they had baseline familiarity, but not necessarily medical fluency, with the target language.  In this case, they used their language ability to evaluate the accuracy of the translation.

\begin{quote}
\textit{I've only really used Google Translate in clinic, in a situation where I'm counseling a patient, and maybe it's on an issue that I just haven't spoken to a patient in Spanish about recently. And so, I just want to make sure, am I using the correct word? I mean, essentially, if I plug it in and I'm looking in Spanish, I can tell whether the translation is correct or not. Whereas in other languages, I can't, which is why I don't use Google Translate for other languages.} (P4, Obstetric and Gynecological Surgeon)
\end{quote}

A few other clinicians noted that some circumstances, including being short on time and having no access to a language translation line, necessitated the use of Google Translate. In such instances, MT was viewed as a last-resort option. 

\begin{quote}
  \textit{Usually, when I'd use it (Google Translate) is out of sheer desperation. So often it was more rare languages where there was no interpreter, we'd be on hold for 15 minutes, realized we were probably not going to get someone at 6:00 AM, while bedside rounding, and just used Google Translate to do the best we could to try to communicate in that setting.} (P19, Family Medicine Physician)
\end{quote}

%\begin{quote}
%\textit{The concerns would be that they're (MT systems) not quite as good as a human translation, because they don't have all of the context. And especially when it comes to video context, the emotional body language cues are really hard for stuff like that to pick up on. But I think it's a tool for quick communication, simple communication. It improves the workflow, it allows you to be more efficient. I think especially for nursing, it would be an awesome tool.} (P13). 
%\end{quote}

Aware that both medical interpreters and MT systems may sometimes cause miscommunication with patients, clinicians had developed techniques to improve communication, and evaluate translation quality and patient understanding, including: \textbf{a) rephrasing medical terms, b) using back translation techniques, c) relying on nonverbal patient cues, and d) testing patient understanding.} When interacting with patients, clinicians often used simple words in their communication. Clinicians frequently employed back translation methods in which they would ask an interpreter to repeat back what had been translated to the patient, often measuring the time the interpreter spent conversing with the patient versus translating the information back. Nonverbal cues from the patient including gestures and facial expressions were often used to gauge the quality of the interpreter. Lastly, teach back methods are a common practice in most training programs (e.g. medical schools), where clinicians are encouraged to ask the patient to repeat back what they said in terms of the patient plan moving forward. Clinicians frequently employed this technique during cross-lingual patient interactions. 

%In our conversations with clinicians, medical interpreters played the role of facilitating cross-lingual communication, but medical literacy rates drove patient understanding and culture differences impacted patient-clinician trust. 

\subsection{Need for Improvements in MT Systems Deployed in Medical Settings}

Clinicians recognized the limitations of machine translation tools. For example, the citing the tools' inability to pick up context like \emph{``emotional body language cues,''} thus making them only suitable for \emph{``quick [and] simple communication.''}

For this reason, clinicians judged the appropriateness of MT on a case-by-case basis, depending on their level of trust in the MT system and the stakes of the interaction. For example, when patient consent was required, clinicians took the time to call a language translation service or a medical interpreter, while in situations they perceived to be lower-stakes, such as taking a patient's medical history, clinicians were more likely to resort to Google Translate to navigate language barriers. 

\begin{quote}
	\textit{I think if it was a more detailed or complex situation, I probably would have called the interpreter. Just again, because legal liability as to whether the patient understood or not.} (P2, Nephrologist)
    \end{quote}

Clinicians in our study also recognized that general purpose machine translation systems such as Google Translate were not built for use in medical settings. One clinician described:

\begin{quote}
	\textit{Even something like Dragon Software, which I don't currently use, but I've used in the past for medical dictation. It's not perfect, so if there's something that's geared towards medical language is important, and that it's been checked out by somebody else} (P4, Obstetric and Gynecological Surgeon)
\end{quote}

Another clinician noted how translation systems often do not account for dialect differences and tone, including local dialects or slangs, cultural differences reflected through tone or inflection, and sarcasm.

%\begin{quote}
%\textit{There's like local dialects, like slang that you can't really figure out. There's cultural differences that you can't really pick up with machines because it's kind of based on tone or inflection, things like that. I don't know how strong your translation services would be in order to account for all of those things. Right? Or like sarcasm or something.} (P16).
%\end{quote}

Thus, lack of trust was directly tied to issues of reliability and accountability. Since translation systems offer little feedback on the quality of their translations, clinicians are advised to use these tools cautiously. However, this guidance is generally vague and \textbf{it is not clear how effectively clinicians are able to exercise caution when using MT.}  Part of this issue is due to the fact that MT tools are not designed specifically for use in healthcare, and evaluations of these systems on medical language suggests a potential for grave harm from translation errors \cite{Vieira21}. As such, many were wary of the tool's accuracy in relation to translating medical language and described accountability concerns, noting that when navigating high-stake medical settings such as soliciting patient consent a lack of trust in the MT system prevented them from relying on these tools. As one physician stated:

 \begin{quote}
   \textit{If it was very complex, where I needed patient's consent, probably would've used a translator, because I knew they would not be supported by Google. Nobody would back me up for the Google Translate.} (P2, Neprologist) \end{quote}

The importance of standardizing evaluation methods of machine translation tools within medical settings was a recurrent theme in our study. Training models on medical phrases can build clinician confidence in MT systems. Clinicians also repeatedly commented on the importance of a machine translation tool having been vetted by a medical society to help them build trust in the MT systems. Medical society validation is crucial because it's typically a signal that the tool was tested via a randomized control trial, and that the findings were published in a peer-reviewed medical journal. 

\begin{quote}
\textit{If it's gone through a vigorous screening process and research and make sure that they're correctly translated and somebody presents to me as a certified medical translation tool, I don't think I would question that.} (P3, Obstetric and Gynecological Surgeon)
\end{quote}

In summary, clinicians recognized some of the limitations of MT systems, but lacked concrete guidelines around when to use these tools in medical settings, or feedback about how they could rephrase terms or otherwise adjust how they use the system to improve translation accuracy.

%\subsection{Implications for Building Trust }

%% file: sections/04_future.tex
Our case study showcased challenges clinicians encountered in their day-to-day practice while using MT and strategies they adopted to try to overcome these challenges. These empirical findings offered new perspectives and opened up conversations on how might we design MT systems for users to build and calibrate trust towards MT, potentially addressing the challenges we mentioned in section \ref{2}. To this end, we suggest that an important first step towards building trustworthy MT systems is to \textbf{conduct empirical studies to understand how people navigate language barriers and whether and how MT systems are being used in specific domains and contexts}. Within the various domains where MT is used (e.g., medicine, law, social media etc.), user needs vary based on the context, for example, the stakes and formality of the interaction, the communication modality, potential cultural and professional gaps, users' language proficiency, and their familiarity with the technology.

The motivation for this approach is three-fold: \emph{First}, understanding the domain-specific needs and challenges of using MT would help researchers and developers \textbf{clarify what user ``trust'' in MT actually means in each specific context}, which is currently understudied in the MT research community. For example, calibrating a clinicians' trust in MT systems in high-stakes medical contexts is different from the level of trust a tourist may need when using an MT system to help them check into a hotel while traveling in a foreign country. In our case study, clinicians' trust towards MT systems was influenced both by their day-to-day interactions with the system and the institutional endorsements based on expert peer-review systems. Other stakeholders under the same context (e.g., patients) and users in other MT use cases might construct and calibrate their trust in drastically different ways based on how often they use MT, whether the system has been rigorously evaluated, and the importance of accurate communication to the task at hand. Prior research has also found that people's trust towards AI systems is shaped by organizational process \cite{yang2019unremarkable, madaio2021assessing}, culture \cite{holstein2019improving}, and power dynamics \cite{Thornton2021FiftySO} when cross-functional actors were collectively interacting with an AI system \cite{rakova2020responsible}. Contextualizing user ``trust'' in MT in real-world practice could advance us a step closer to providing nuanced, situated indicators of translation quality (Challenge \ref{2.2}), and understanding how to prevent or help users identify and cope with rare but high impact errors (Challenge  \ref{2.3})

\emph{Second}, researchers and developers could \textbf{design interfaces and interactions that build upon MT users' current practice and strategies surfaced from the empirical research}. HCI has a long tradition of drawing insight and inspiration from users to design ecologically valid products \cite{elizabeth2002harnessing}. Our semi-structured interviews, for example, surfaced that clinicians currently use both back translation techniques and ``teach back methods'' in their daily communications with patients mediated by human interpreters. Future designs could leverage and incorporate the interactive and reciprocal strategies that clinicians have already developed to improve communication with patients (Challenge \ref{2.1}, \ref{2.4}). Besides semi-structured interviews, other methods like participatory design \cite{Kuhn1993ParticipatoryD}, diary studies \cite{rieman1993diary}, value-sensitive design \cite{friedman1996value}, and speed dating \cite{zimmerman2017speed} (and the combination of these methods) all have the potential to collect useful data in different formats to help designer future MT systems based on stakeholders' needs (Challenge \ref{2.4}).

It is also possible that, through analyzing and making sense of fieldwork data, researchers could surface the latent needs and unspoken desires reflected through users' current practice and interaction patterns, and thus inform future design \cite{William2001HarnessingPC}. For example, there was general acknowledgment among clinicians that MT tools should be given to patients so they can communicate their symptoms to clinicians instead of always having clinicians holding the MT tools. Having patients type their symptoms and medical questions into a tool on their own could yield a more comprehensive medical history. This observation leads us to the design implication that users' \emph{personal agency} while using MT plays an important role in trust-building. 

Third, studying MT systems' use in specialized contexts could help researchers prioritize \textbf{designing and developing domain-specific, context-aware MT systems.} Current commercial MT systems are mostly general-purpose \cite{Wu2016GooglesNM}, but trustworthy MT systems can be specifically designed to prioritize accurate translation of the specific language used under the specific context, possibly through domain adaptation methods \cite{Finn2017ModelAgnosticMF}. Unfortunately, domain-specific model-building and quality evaluation for MT systems are currently understudied. Field data from users could potentially provide useful training datasets and instantiate theoretical frameworks like context-aware learning for neural machine translation \cite{jean2019context}, narrowing down the design space for MT (Challenge \ref{2.5}). MT researchers could also draw from prior HCI research in context-aware ubiquitous computing \cite{schilit1994context} and mixed-initiative interfaces \cite{allen1999mixed} to design MT tools that are beyond text-based input-output box translation interfaces (Challenge \ref{2.1}). Another promising future direction that could complement current progress in neural machine translation is leveraging relatively standardized patterns and topics of conversation to introduce fixed phrase-based translations \cite{turner2019evaluating, spechbach2017comparison, panayiotou2020perceptions}(Challenge \ref{2.5})
%For example, using medical terminologies, discussing medical history, and describing current symptoms are all common

In addition to building domain-specific models, it is also crucial to draw from empirical findings to create both general and domain-specific user guidelines \cite{madaio2020co, googlePAIR}, training and onboarding materials \cite{cai2019hello, brundage2020toward} for MT users, and rigorous evaluation and endorsement \cite{Khoong2019AssessingTU, Spechbach2017ComparisonOT, Turner2019EvaluatingTU}. In short, we see value in building domain-specific MT ecosystems.
\section{Conclusion}

In this position paper, we first synthesize challenges in building trustworthy Machine Translation systems. Then, we share a case study in which we conducted semi-structured interviews with 20 clinicians to understand how MT was used in their daily practices and their needs for trust in medical MT systems. As a conclusion, we call for conducting empirical studies to understand how people communicate across languages and how MT systems are being used under specific contexts as an important but often neglected step towards building trustworthy MT systems. We explained how this approach could potentially address the unique challenges in building trustworthy MT, followed by design implications that are generalizable to MT systems for contexts beyond medical settings.

%% file: 0_main.bbl

\begin{thebibliography}{67}

%%% the Web address.

\ifx \showCODEN    \undefined \def \showCODEN     #1{\unskip}     \fi
\ifx \showDOI      \undefined \def \showDOI       #1{#1}\fi
\ifx \showISBNx    \undefined \def \showISBNx     #1{\unskip}     \fi
\ifx \showISBNxiii \undefined \def \showISBNxiii  #1{\unskip}     \fi
\ifx \showISSN     \undefined \def \showISSN      #1{\unskip}     \fi
\ifx \showLCCN     \undefined \def \showLCCN      #1{\unskip}     \fi
\ifx \shownote     \undefined \def \shownote      #1{#1}          \fi
\ifx \showarticletitle \undefined \def \showarticletitle #1{#1}   \fi
\ifx \showURL      \undefined \def \showURL       {\relax}        \fi
% invisible to TeX
\providecommand\bibfield[2]{#2}
\providecommand\bibinfo[2]{#2}
\providecommand\natexlab[1]{#1}
\providecommand\showeprint[2][]{arXiv:#2}

\bibitem[\protect\citeauthoryear{??}{goo}{2021}]%
        {googlePAIR}
 \bibinfo{year}{2021}\natexlab{}.
\newblock \showarticletitle{People AI Guidebook}.
\newblock  (\bibinfo{year}{2021}).
\newblock
\urldef\tempurl%
\url{https://pair.withgoogle.com/guidebook/}
\showURL{%
\tempurl}


\bibitem[\protect\citeauthoryear{Allen, Guinn, and Horvtz}{Allen
  et~al\mbox{.}}{1999}]%
        {allen1999mixed}
\bibfield{author}{\bibinfo{person}{James~E Allen}, \bibinfo{person}{Curry~I
  Guinn}, {and} \bibinfo{person}{Eric Horvtz}.}
  \bibinfo{year}{1999}\natexlab{}.
\newblock \showarticletitle{Mixed-initiative interaction}.
\newblock \bibinfo{journal}{\emph{IEEE Intelligent Systems and their
  Applications}} \bibinfo{volume}{14}, \bibinfo{number}{5}
  (\bibinfo{year}{1999}), \bibinfo{pages}{14--23}.
\newblock


\bibitem[\protect\citeauthoryear{Bahdanau, Cho, and Bengio}{Bahdanau
  et~al\mbox{.}}{2015}]%
        {Bahdanau2015NeuralMT}
\bibfield{author}{\bibinfo{person}{Dzmitry Bahdanau},
  \bibinfo{person}{Kyunghyun Cho}, {and} \bibinfo{person}{Yoshua Bengio}.}
  \bibinfo{year}{2015}\natexlab{}.
\newblock \showarticletitle{Neural Machine Translation by Jointly Learning to
  Align and Translate}.
\newblock \bibinfo{journal}{\emph{CoRR}}  \bibinfo{volume}{abs/1409.0473}
  (\bibinfo{year}{2015}).
\newblock


\bibitem[\protect\citeauthoryear{Bender, Gebru, McMillan-Major, and
  Shmitchell}{Bender et~al\mbox{.}}{2021}]%
        {bender2021dangers}
\bibfield{author}{\bibinfo{person}{Emily~M Bender}, \bibinfo{person}{Timnit
  Gebru}, \bibinfo{person}{Angelina McMillan-Major}, {and}
  \bibinfo{person}{Shmargaret Shmitchell}.} \bibinfo{year}{2021}\natexlab{}.
\newblock \showarticletitle{On the Dangers of Stochastic Parrots: Can Language
  Models Be Too Big?}. In \bibinfo{booktitle}{\emph{Proceedings of the 2021 ACM
  Conference on Fairness, Accountability, and Transparency}}.
  \bibinfo{pages}{610--623}.
\newblock


\bibitem[\protect\citeauthoryear{Berger}{Berger}{2017}]%
        {berger2017attack}
\bibfield{author}{\bibinfo{person}{Yotam Berger}.}
  \bibinfo{year}{2017}\natexlab{}.
\newblock \showarticletitle{Israel Arrests Palestinian Because Facebook
  Translated 'Good Morning' to 'Attack Them'}.
\newblock \bibinfo{journal}{\emph{Haaretz}} (\bibinfo{date}{Oct}
  \bibinfo{year}{2017}).
\newblock
\urldef\tempurl%
\url{https://www.haaretz.com/israel-news/palestinian-arrested-over-mistranslated-good-morning-facebook-post-1.5459427}
\showURL{%
\tempurl}


\bibitem[\protect\citeauthoryear{Brundage, Avin, Wang, Belfield, Krueger,
  Hadfield, Khlaaf, Yang, Toner, Fong, et~al\mbox{.}}{Brundage
  et~al\mbox{.}}{2020}]%
        {brundage2020toward}
\bibfield{author}{\bibinfo{person}{Miles Brundage}, \bibinfo{person}{Shahar
  Avin}, \bibinfo{person}{Jasmine Wang}, \bibinfo{person}{Haydn Belfield},
  \bibinfo{person}{Gretchen Krueger}, \bibinfo{person}{Gillian Hadfield},
  \bibinfo{person}{Heidy Khlaaf}, \bibinfo{person}{Jingying Yang},
  \bibinfo{person}{Helen Toner}, \bibinfo{person}{Ruth Fong}, {et~al\mbox{.}}}
  \bibinfo{year}{2020}\natexlab{}.
\newblock \showarticletitle{Toward trustworthy AI development: mechanisms for
  supporting verifiable claims}.
\newblock \bibinfo{journal}{\emph{arXiv preprint arXiv:2004.07213}}
  (\bibinfo{year}{2020}).
\newblock


\bibitem[\protect\citeauthoryear{Cai, Winter, Steiner, Wilcox, and Terry}{Cai
  et~al\mbox{.}}{2019}]%
        {cai2019hello}
\bibfield{author}{\bibinfo{person}{Carrie~J Cai}, \bibinfo{person}{Samantha
  Winter}, \bibinfo{person}{David Steiner}, \bibinfo{person}{Lauren Wilcox},
  {and} \bibinfo{person}{Michael Terry}.} \bibinfo{year}{2019}\natexlab{}.
\newblock \showarticletitle{" Hello AI": uncovering the onboarding needs of
  medical practitioners for human-AI collaborative decision-making}.
\newblock \bibinfo{journal}{\emph{Proceedings of the ACM on Human-computer
  Interaction}} \bibinfo{volume}{3}, \bibinfo{number}{CSCW}
  (\bibinfo{year}{2019}), \bibinfo{pages}{1--24}.
\newblock


\bibitem[\protect\citeauthoryear{Center}{Center}{[n.d.]a}]%
        {facebooktranslate}
\bibfield{author}{\bibinfo{person}{Facebook~Help Center}.}
  \bibinfo{year}{[n.d.]}\natexlab{a}.
\newblock \bibinfo{booktitle}{\emph{Translate Facebook APP}}.
\newblock
\urldef\tempurl%
\url{https://www.facebook.com/help/364760868183100}
\showURL{%
\tempurl}


\bibitem[\protect\citeauthoryear{Center}{Center}{[n.d.]b}]%
        {twittertranslate}
\bibfield{author}{\bibinfo{person}{Twitter~Help Center}.}
  \bibinfo{year}{[n.d.]}\natexlab{b}.
\newblock \bibinfo{booktitle}{\emph{Tweet Translation}}.
\newblock
\urldef\tempurl%
\url{https://help.twitter.com/en/using-twitter/translate-tweets}
\showURL{%
\tempurl}


\bibitem[\protect\citeauthoryear{Das, Kuznetsova, Zhu, and Milanaik}{Das
  et~al\mbox{.}}{2019}]%
        {das2019dangers}
\bibfield{author}{\bibinfo{person}{Prithwijit Das}, \bibinfo{person}{Anna
  Kuznetsova}, \bibinfo{person}{Meng'ou Zhu}, {and} \bibinfo{person}{Ruth
  Milanaik}.} \bibinfo{year}{2019}\natexlab{}.
\newblock \showarticletitle{Dangers of Machine Translation: The Need for
  Professionally Translated Anticipatory Guidance Resources for Limited English
  Proficiency Caregivers}.
\newblock \bibinfo{journal}{\emph{Clinical Pediatrics (Phila)}}
  \bibinfo{volume}{58}, \bibinfo{number}{2} (\bibinfo{date}{Feb}
  \bibinfo{year}{2019}), \bibinfo{pages}{247--249}.
\newblock
\urldef\tempurl%
\url{https://doi.org/10.1177/0009922818809494}
\showDOI{\tempurl}


\bibitem[\protect\citeauthoryear{Elizabeth and William}{Elizabeth and
  William}{2002}]%
        {elizabeth2002harnessing}
\bibfield{author}{\bibinfo{person}{B-N~Sanders Elizabeth} {and}
  \bibinfo{person}{Colin~T William}.} \bibinfo{year}{2002}\natexlab{}.
\newblock \showarticletitle{Harnessing people’s creativity: Ideation and
  expression through visual communication}.
\newblock \bibinfo{journal}{\emph{Focus groups: Supporting effective product
  development}}  \bibinfo{volume}{137} (\bibinfo{year}{2002}).
\newblock


\bibitem[\protect\citeauthoryear{Finn, Abbeel, and Levine}{Finn
  et~al\mbox{.}}{2017}]%
        {Finn2017ModelAgnosticMF}
\bibfield{author}{\bibinfo{person}{Chelsea Finn}, \bibinfo{person}{P. Abbeel},
  {and} \bibinfo{person}{Sergey Levine}.} \bibinfo{year}{2017}\natexlab{}.
\newblock \showarticletitle{Model-Agnostic Meta-Learning for Fast Adaptation of
  Deep Networks}. In \bibinfo{booktitle}{\emph{ICML}}.
\newblock


\bibitem[\protect\citeauthoryear{Friedman}{Friedman}{1996}]%
        {friedman1996value}
\bibfield{author}{\bibinfo{person}{Batya Friedman}.}
  \bibinfo{year}{1996}\natexlab{}.
\newblock \showarticletitle{Value-sensitive design}.
\newblock \bibinfo{journal}{\emph{interactions}} \bibinfo{volume}{3},
  \bibinfo{number}{6} (\bibinfo{year}{1996}), \bibinfo{pages}{16--23}.
\newblock


\bibitem[\protect\citeauthoryear{Gao, Wang, Cosley, and Fussell}{Gao
  et~al\mbox{.}}{2013}]%
        {gao2013highlighting}
\bibfield{author}{\bibinfo{person}{Ge Gao}, \bibinfo{person}{Hao-Chuan Wang},
  \bibinfo{person}{Dan Cosley}, {and} \bibinfo{person}{Susan~R. Fussell}.}
  \bibinfo{year}{2013}\natexlab{}.
\newblock \showarticletitle{Same translation but different experience: the
  effects of highlighting on machine-translated conversations}. In
  \bibinfo{booktitle}{\emph{Proceedings of the {SIGCHI} {Conference} on {Human}
  {Factors} in {Computing} {Systems}}} \emph{(\bibinfo{series}{{CHI} '13})}.
  \bibinfo{publisher}{Association for Computing Machinery},
  \bibinfo{address}{Paris, France}, \bibinfo{pages}{449--458}.
\newblock
\showISBNx{978-1-4503-1899-0}
\urldef\tempurl%
\url{https://doi.org/10.1145/2470654.2470719}
\showDOI{\tempurl}


\bibitem[\protect\citeauthoryear{Gao, Xu, Hau, Yao, Cosley, and Fussell}{Gao
  et~al\mbox{.}}{2015}]%
        {gao2015twoOutputs}
\bibfield{author}{\bibinfo{person}{Ge Gao}, \bibinfo{person}{Bin Xu},
  \bibinfo{person}{David~C. Hau}, \bibinfo{person}{Zheng Yao},
  \bibinfo{person}{Dan Cosley}, {and} \bibinfo{person}{Susan~R. Fussell}.}
  \bibinfo{year}{2015}\natexlab{}.
\newblock \showarticletitle{Two is {Better} {Than} {One}: {Improving}
  {Multilingual} {Collaboration} by {Giving} {Two} {Machine} {Translation}
  {Outputs}}. In \bibinfo{booktitle}{\emph{Proceedings of the 18th {ACM}
  {Conference} on {Computer} {Supported} {Cooperative} {Work} \& {Social}
  {Computing} - {CSCW} '15}}. \bibinfo{publisher}{ACM Press},
  \bibinfo{address}{Vancouver, BC, Canada}, \bibinfo{pages}{852--863}.
\newblock
\showISBNx{978-1-4503-2922-4}
\urldef\tempurl%
\url{https://doi.org/10.1145/2675133.2675197}
\showDOI{\tempurl}


\bibitem[\protect\citeauthoryear{Gille, Jobin, and Ienca}{Gille
  et~al\mbox{.}}{2020}]%
        {gille2020we}
\bibfield{author}{\bibinfo{person}{Felix Gille}, \bibinfo{person}{Anna Jobin},
  {and} \bibinfo{person}{Marcello Ienca}.} \bibinfo{year}{2020}\natexlab{}.
\newblock \showarticletitle{What we talk about when we talk about trust: theory
  of trust for AI in healthcare}.
\newblock \bibinfo{journal}{\emph{Intelligence-Based Medicine}}
  \bibinfo{volume}{1} (\bibinfo{year}{2020}), \bibinfo{pages}{100001}.
\newblock


\bibitem[\protect\citeauthoryear{Holstein, Wortman~Vaughan, Daum{\'e}~III,
  Dudik, and Wallach}{Holstein et~al\mbox{.}}{2019}]%
        {holstein2019improving}
\bibfield{author}{\bibinfo{person}{Kenneth Holstein}, \bibinfo{person}{Jennifer
  Wortman~Vaughan}, \bibinfo{person}{Hal Daum{\'e}~III}, \bibinfo{person}{Miro
  Dudik}, {and} \bibinfo{person}{Hanna Wallach}.}
  \bibinfo{year}{2019}\natexlab{}.
\newblock \showarticletitle{Improving fairness in machine learning systems:
  What do industry practitioners need?}. In
  \bibinfo{booktitle}{\emph{Proceedings of the 2019 CHI Conference on Human
  Factors in Computing Systems}}. \bibinfo{pages}{1--16}.
\newblock


\bibitem[\protect\citeauthoryear{Hsu, Huang, Verma, Mauri, Nourbakhsh, and
  Bozzon}{Hsu et~al\mbox{.}}{2021}]%
        {hsu2021empowering}
\bibfield{author}{\bibinfo{person}{Yen-Chia Hsu},
  \bibinfo{person}{Ting-Hao'Kenneth' Huang}, \bibinfo{person}{Himanshu Verma},
  \bibinfo{person}{Andrea Mauri}, \bibinfo{person}{Illah Nourbakhsh}, {and}
  \bibinfo{person}{Alessandro Bozzon}.} \bibinfo{year}{2021}\natexlab{}.
\newblock \showarticletitle{Empowering Local Communities Using Artificial
  Intelligence}.
\newblock \bibinfo{journal}{\emph{arXiv preprint arXiv:2110.02007}}
  (\bibinfo{year}{2021}).
\newblock


\bibitem[\protect\citeauthoryear{Jean and Cho}{Jean and Cho}{2019}]%
        {jean2019context}
\bibfield{author}{\bibinfo{person}{S{\'e}bastien Jean} {and}
  \bibinfo{person}{Kyunghyun Cho}.} \bibinfo{year}{2019}\natexlab{}.
\newblock \showarticletitle{Context-aware learning for neural machine
  translation}.
\newblock \bibinfo{journal}{\emph{arXiv preprint arXiv:1903.04715}}
  (\bibinfo{year}{2019}).
\newblock


\bibitem[\protect\citeauthoryear{Kaur, Nori, Jenkins, Caruana, Wallach, and
  Wortman~Vaughan}{Kaur et~al\mbox{.}}{2020}]%
        {kaur2020interpreting}
\bibfield{author}{\bibinfo{person}{Harmanpreet Kaur}, \bibinfo{person}{Harsha
  Nori}, \bibinfo{person}{Samuel Jenkins}, \bibinfo{person}{Rich Caruana},
  \bibinfo{person}{Hanna Wallach}, {and} \bibinfo{person}{Jennifer
  Wortman~Vaughan}.} \bibinfo{year}{2020}\natexlab{}.
\newblock \bibinfo{booktitle}{\emph{Interpreting Interpretability:
  Understanding Data Scientists' Use of Interpretability Tools for Machine
  Learning}}.
\newblock \bibinfo{publisher}{Association for Computing Machinery},
  \bibinfo{address}{New York, NY, USA}, \bibinfo{pages}{1–14}.
\newblock
\showISBNx{9781450367080}
\urldef\tempurl%
\url{https://doi.org/10.1145/3313831.3376219}
\showURL{%
\tempurl}


\bibitem[\protect\citeauthoryear{Khoong and Rodriguez}{Khoong and
  Rodriguez}{2022}]%
        {khoong2022research}
\bibfield{author}{\bibinfo{person}{Elaine~C Khoong} {and}
  \bibinfo{person}{Jorge~A Rodriguez}.} \bibinfo{year}{2022}\natexlab{}.
\newblock \showarticletitle{A Research Agenda for Using Machine Translation in
  Clinical Medicine}.
\newblock \bibinfo{journal}{\emph{Journal of General Internal Medicine}}
  (\bibinfo{year}{2022}), \bibinfo{pages}{1--3}.
\newblock


\bibitem[\protect\citeauthoryear{Khoong, Steinbrook, Brown, and
  Fernandez}{Khoong et~al\mbox{.}}{2019a}]%
        {Khoong19}
\bibfield{author}{\bibinfo{person}{Elaine~C Khoong}, \bibinfo{person}{Eric
  Steinbrook}, \bibinfo{person}{Cortlyn Brown}, {and} \bibinfo{person}{Alicia
  Fernandez}.} \bibinfo{year}{2019}\natexlab{a}.
\newblock \showarticletitle{Assessing the use of Google Translate for Spanish
  and Chinese translations of emergency department discharge instructions}.
\newblock \bibinfo{journal}{\emph{JAMA internal medicine}}
  \bibinfo{volume}{179}, \bibinfo{number}{4} (\bibinfo{year}{2019}),
  \bibinfo{pages}{580--582}.
\newblock


\bibitem[\protect\citeauthoryear{Khoong, Steinbrook, Brown, and
  Fern{\'a}ndez}{Khoong et~al\mbox{.}}{2019b}]%
        {Khoong2019AssessingTU}
\bibfield{author}{\bibinfo{person}{Elaine~C. Khoong}, \bibinfo{person}{Eric
  Steinbrook}, \bibinfo{person}{Cortlyn Brown}, {and} \bibinfo{person}{Alicia
  Fern{\'a}ndez}.} \bibinfo{year}{2019}\natexlab{b}.
\newblock \showarticletitle{Assessing the Use of Google Translate for Spanish
  and Chinese Translations of Emergency Department Discharge Instructions}.
\newblock \bibinfo{journal}{\emph{JAMA Internal Medicine}}
  \bibinfo{volume}{179} (\bibinfo{year}{2019}), \bibinfo{pages}{580–582}.
\newblock


\bibitem[\protect\citeauthoryear{Kuhn and Muller}{Kuhn and Muller}{1993}]%
        {Kuhn1993ParticipatoryD}
\bibfield{author}{\bibinfo{person}{Sarah Kuhn} {and}
  \bibinfo{person}{Michael~J. Muller}.} \bibinfo{year}{1993}\natexlab{}.
\newblock \showarticletitle{Participatory design}.
\newblock \bibinfo{journal}{\emph{Commun. ACM}}  \bibinfo{volume}{36}
  (\bibinfo{year}{1993}), \bibinfo{pages}{24--28}.
\newblock


\bibitem[\protect\citeauthoryear{Liebling, Lahav, Evans, Donsbach, Holbrook,
  Smus, and Boran}{Liebling et~al\mbox{.}}{2020}]%
        {liebling2020unmetneeds}
\bibfield{author}{\bibinfo{person}{Daniel~J. Liebling}, \bibinfo{person}{Michal
  Lahav}, \bibinfo{person}{Abigail Evans}, \bibinfo{person}{Aaron Donsbach},
  \bibinfo{person}{Jess Holbrook}, \bibinfo{person}{Boris Smus}, {and}
  \bibinfo{person}{Lindsey Boran}.} \bibinfo{year}{2020}\natexlab{}.
\newblock \showarticletitle{Unmet Needs and Opportunities for Mobile
  Translation {AI}}. In \bibinfo{booktitle}{\emph{Proceedings of the 2020 CHI
  Conference on Human Factors in Computing Systems}}.
  \bibinfo{publisher}{Association for Computing Machinery},
  \bibinfo{address}{New York, NY, USA}, \bibinfo{pages}{1–13}.
\newblock
\showISBNx{9781450367080}
\urldef\tempurl%
\url{https://doi.org/10.1145/3313831.3376261}
\showURL{%
\tempurl}


\bibitem[\protect\citeauthoryear{Lim, Cosley, and Fussell}{Lim
  et~al\mbox{.}}{2018}]%
        {lim2018sensetrans}
\bibfield{author}{\bibinfo{person}{Hajin Lim}, \bibinfo{person}{Dan Cosley},
  {and} \bibinfo{person}{Susan~R. Fussell}.} \bibinfo{year}{2018}\natexlab{}.
\newblock \showarticletitle{Beyond {Translation}: {Design} and {Evaluation} of
  an {Emotional} and {Contextual} {Knowledge} {Interface} for {Foreign}
  {Language} {Social} {Media} {Posts}}. In
  \bibinfo{booktitle}{\emph{Proceedings of the 2018 {CHI} {Conference} on
  {Human} {Factors} in {Computing} {Systems}}} \emph{(\bibinfo{series}{{CHI}
  '18})}. \bibinfo{publisher}{Association for Computing Machinery},
  \bibinfo{address}{Montreal QC, Canada}, \bibinfo{pages}{1--12}.
\newblock
\showISBNx{978-1-4503-5620-6}
\urldef\tempurl%
\url{https://doi.org/10.1145/3173574.3173791}
\showDOI{\tempurl}


\bibitem[\protect\citeauthoryear{Liu, Faes, Kale, Wagner, Fu, Bruynseels,
  Mahendiran, Moraes, Shamdas, Kern, Ledsam, Schmid, Balaskas, Topol, Bachmann,
  Keane, and Denniston}{Liu et~al\mbox{.}}{2019}]%
        {liu2019comparison}
\bibfield{author}{\bibinfo{person}{Xiaoxuan Liu}, \bibinfo{person}{Livia Faes},
  \bibinfo{person}{Aditya~U Kale}, \bibinfo{person}{Siegfried~K Wagner},
  \bibinfo{person}{Dun~Jack Fu}, \bibinfo{person}{Alice Bruynseels},
  \bibinfo{person}{Thushika Mahendiran}, \bibinfo{person}{Gabriella Moraes},
  \bibinfo{person}{Mohith Shamdas}, \bibinfo{person}{Christoph Kern},
  \bibinfo{person}{Joseph~R Ledsam}, \bibinfo{person}{Martin~K Schmid},
  \bibinfo{person}{Konstantinos Balaskas}, \bibinfo{person}{Eric~J Topol},
  \bibinfo{person}{Lucas~M Bachmann}, \bibinfo{person}{Pearse~A Keane}, {and}
  \bibinfo{person}{Alastair~K Denniston}.} \bibinfo{year}{2019}\natexlab{}.
\newblock \showarticletitle{A comparison of deep learning performance against
  health-care professionals in detecting diseases from medical imaging: a
  systematic review and meta-analysis}.
\newblock \bibinfo{journal}{\emph{The Lancet Digital Health}}
  \bibinfo{volume}{1}, \bibinfo{number}{6} (\bibinfo{year}{2019}).
\newblock
\urldef\tempurl%
\url{https://doi.org/10.1016/S2589-7500(19)30123-2}
\showDOI{\tempurl}


\bibitem[\protect\citeauthoryear{Madaio, Egede, Subramonyam, Vaughan, and
  Wallach}{Madaio et~al\mbox{.}}{2021}]%
        {madaio2021assessing}
\bibfield{author}{\bibinfo{person}{Michael Madaio}, \bibinfo{person}{Lisa
  Egede}, \bibinfo{person}{Hariharan Subramonyam},
  \bibinfo{person}{Jennifer~Wortman Vaughan}, {and} \bibinfo{person}{Hanna
  Wallach}.} \bibinfo{year}{2021}\natexlab{}.
\newblock \showarticletitle{Assessing the Fairness of AI Systems: AI
  Practitioners' Processes, Challenges, and Needs for Support}.
\newblock \bibinfo{journal}{\emph{arXiv preprint arXiv:2112.05675}}
  (\bibinfo{year}{2021}).
\newblock


\bibitem[\protect\citeauthoryear{Madaio, Stark, Wortman~Vaughan, and
  Wallach}{Madaio et~al\mbox{.}}{2020}]%
        {madaio2020co}
\bibfield{author}{\bibinfo{person}{Michael~A Madaio}, \bibinfo{person}{Luke
  Stark}, \bibinfo{person}{Jennifer Wortman~Vaughan}, {and}
  \bibinfo{person}{Hanna Wallach}.} \bibinfo{year}{2020}\natexlab{}.
\newblock \showarticletitle{Co-designing checklists to understand
  organizational challenges and opportunities around fairness in ai}. In
  \bibinfo{booktitle}{\emph{Proceedings of the 2020 CHI Conference on Human
  Factors in Computing Systems}}. \bibinfo{pages}{1--14}.
\newblock


\bibitem[\protect\citeauthoryear{Martindale and Carpuat}{Martindale and
  Carpuat}{2018}]%
        {martindale_fluency_2018}
\bibfield{author}{\bibinfo{person}{Marianna Martindale} {and}
  \bibinfo{person}{Marine Carpuat}.} \bibinfo{year}{2018}\natexlab{}.
\newblock \showarticletitle{Fluency {Over} {Adequacy}: {A} {Pilot} {Study} in
  {Measuring} {User} {Trust} in {Imperfect} {MT}}. In
  \bibinfo{booktitle}{\emph{Proceedings of the 13th {Conference} of the
  {Association} for {Machine} {Translation} in the {Americas} ({Volume} 1:
  {Research} {Papers})}}. \bibinfo{publisher}{Association for Machine
  Translation in the Americas}, \bibinfo{address}{Boston, MA, USA},
  \bibinfo{pages}{13--25}.
\newblock
\urldef\tempurl%
\url{https://www.aclweb.org/anthology/W18-1803}
\showURL{%
\tempurl}


\bibitem[\protect\citeauthoryear{Mehandru, Robertson, and Salehi}{Mehandru
  et~al\mbox{.}}{2022}]%
        {mehandru_2022_reliable}
\bibfield{author}{\bibinfo{person}{Nikita Mehandru}, \bibinfo{person}{Samantha
  Robertson}, {and} \bibinfo{person}{Niloufar Salehi}.}
  \bibinfo{year}{2022}\natexlab{}.
\newblock \showarticletitle{Reliable and Safe Use Machine Translation in
  Medical Settings}. In \bibinfo{booktitle}{\emph{Proceedings of the 2022 ACM
  Conference on Fairness, Accountability, and Transparency}} (Seoul, South
  Korea) \emph{(\bibinfo{series}{FAccT '22})}. \bibinfo{publisher}{Association
  for Computing Machinery}, \bibinfo{address}{New York, NY, USA}.
\newblock


\bibitem[\protect\citeauthoryear{Miyabe and Yoshino}{Miyabe and
  Yoshino}{2009}]%
        {miyabe2009backtranslation}
\bibfield{author}{\bibinfo{person}{Mai Miyabe} {and} \bibinfo{person}{Takashi
  Yoshino}.} \bibinfo{year}{2009}\natexlab{}.
\newblock \showarticletitle{Accuracy Evaluation of Sentences Translated to
  Intermediate Language in Back Translation}. In
  \bibinfo{booktitle}{\emph{Proceedings of the 3rd International Universal
  Communication Symposium}} (Tokyo, Japan) \emph{(\bibinfo{series}{IUCS '09})}.
  \bibinfo{publisher}{Association for Computing Machinery},
  \bibinfo{address}{New York, NY, USA}, \bibinfo{pages}{30–35}.
\newblock
\showISBNx{9781605586410}
\urldef\tempurl%
\url{https://doi.org/10.1145/1667780.1667787}
\showDOI{\tempurl}


\bibitem[\protect\citeauthoryear{Miyabe and Yoshino}{Miyabe and
  Yoshino}{2010}]%
        {miyabe2010detecting}
\bibfield{author}{\bibinfo{person}{Mai Miyabe} {and} \bibinfo{person}{Takashi
  Yoshino}.} \bibinfo{year}{2010}\natexlab{}.
\newblock \showarticletitle{Influence of Detecting Inaccurate Messages in
  Real-Time Remote Text-Based Communication via Machine Translation}. In
  \bibinfo{booktitle}{\emph{Proceedings of the 3rd International Conference on
  Intercultural Collaboration}} (Copenhagen, Denmark)
  \emph{(\bibinfo{series}{ICIC '10})}. \bibinfo{publisher}{Association for
  Computing Machinery}, \bibinfo{address}{New York, NY, USA},
  \bibinfo{pages}{59–68}.
\newblock
\showISBNx{9781450301084}
\urldef\tempurl%
\url{https://doi.org/10.1145/1841853.1841863}
\showDOI{\tempurl}


\bibitem[\protect\citeauthoryear{Miyabe and Yoshino}{Miyabe and
  Yoshino}{2011}]%
        {miyabe2011can}
\bibfield{author}{\bibinfo{person}{Mai Miyabe} {and} \bibinfo{person}{Takashi
  Yoshino}.} \bibinfo{year}{2011}\natexlab{}.
\newblock \showarticletitle{Can Indicating Translation Accuracy Encourage
  People to Rectify Inaccurate Translations?}. In
  \bibinfo{booktitle}{\emph{Human-Computer Interaction. Interaction Techniques
  and Environments}}, \bibfield{editor}{\bibinfo{person}{Julie~A. Jacko}}
  (Ed.). \bibinfo{publisher}{Springer Berlin Heidelberg},
  \bibinfo{address}{Berlin, Heidelberg}, \bibinfo{pages}{368--377}.
\newblock
\showISBNx{978-3-642-21605-3}


\bibitem[\protect\citeauthoryear{Moberly}{Moberly}{2018}]%
        {moberly2018choose}
\bibfield{author}{\bibinfo{person}{Tom Moberly}.}
  \bibinfo{year}{2018}\natexlab{}.
\newblock \showarticletitle{Doctors choose Google Translate to communicate with
  patients because of easy access}.
\newblock \bibinfo{journal}{\emph{BMJ}}  \bibinfo{volume}{362}
  (\bibinfo{year}{2018}).
\newblock
\showISSN{0959-8138}
\urldef\tempurl%
\url{https://doi.org/10.1136/bmj.k3974}
\showDOI{\tempurl}
\showeprint{https://www.bmj.com/content/362/bmj.k3974.full.pdf}


\bibitem[\protect\citeauthoryear{Panayiotou, Hwang, Williams, Chong, LoGiudice,
  Haralambous, Lin, Zucchi, Mascitti-Meuter, Goh, You, and
  Batchelor}{Panayiotou et~al\mbox{.}}{2020}]%
        {panayiotou2020perceptions}
\bibfield{author}{\bibinfo{person}{Anita Panayiotou}, \bibinfo{person}{Kerry
  Hwang}, \bibinfo{person}{Sue Williams}, \bibinfo{person}{Terence W~H Chong},
  \bibinfo{person}{Dina LoGiudice}, \bibinfo{person}{Betty Haralambous},
  \bibinfo{person}{Xiaoping Lin}, \bibinfo{person}{Emiliano Zucchi},
  \bibinfo{person}{Monita Mascitti-Meuter}, \bibinfo{person}{Anita M~Y Goh},
  \bibinfo{person}{Emily You}, {and} \bibinfo{person}{Frances Batchelor}.}
  \bibinfo{year}{2020}\natexlab{}.
\newblock \showarticletitle{The perceptions of translation apps for everyday
  health care in healthcare workers and older people: A multi-method study}.
\newblock \bibinfo{journal}{\emph{Journal of Clinical Nursing}}
  \bibinfo{volume}{29}, \bibinfo{number}{17-18} (\bibinfo{date}{Sep}
  \bibinfo{year}{2020}), \bibinfo{pages}{3516--3526}.
\newblock
\urldef\tempurl%
\url{https://doi.org/10.1111/jocn.15390}
\showDOI{\tempurl}


\bibitem[\protect\citeauthoryear{Papineni, Roukos, Ward, and Zhu}{Papineni
  et~al\mbox{.}}{2002}]%
        {Papineni2002BleuAM}
\bibfield{author}{\bibinfo{person}{Kishore Papineni}, \bibinfo{person}{Salim
  Roukos}, \bibinfo{person}{Todd Ward}, {and} \bibinfo{person}{Wei-Jing Zhu}.}
  \bibinfo{year}{2002}\natexlab{}.
\newblock \showarticletitle{Bleu: a Method for Automatic Evaluation of Machine
  Translation}. In \bibinfo{booktitle}{\emph{ACL}}.
\newblock


\bibitem[\protect\citeauthoryear{Pierre, Crooks, Currie, Paris, and
  Pasquetto}{Pierre et~al\mbox{.}}{2021}]%
        {pierre2021getting}
\bibfield{author}{\bibinfo{person}{Jennifer Pierre}, \bibinfo{person}{Roderic
  Crooks}, \bibinfo{person}{Morgan Currie}, \bibinfo{person}{Britt Paris},
  {and} \bibinfo{person}{Irene Pasquetto}.} \bibinfo{year}{2021}\natexlab{}.
\newblock \showarticletitle{Getting Ourselves Together: Data-centered
  participatory design research \& epistemic burden}. In
  \bibinfo{booktitle}{\emph{Proceedings of the 2021 CHI Conference on Human
  Factors in Computing Systems}}. \bibinfo{pages}{1--11}.
\newblock


\bibitem[\protect\citeauthoryear{Post}{Post}{2018}]%
        {Post2018ACF}
\bibfield{author}{\bibinfo{person}{Matt Post}.}
  \bibinfo{year}{2018}\natexlab{}.
\newblock \showarticletitle{A Call for Clarity in Reporting BLEU Scores}. In
  \bibinfo{booktitle}{\emph{WMT}}.
\newblock


\bibitem[\protect\citeauthoryear{Rakova, Yang, Cramer, and Chowdhury}{Rakova
  et~al\mbox{.}}{2020}]%
        {rakova2020responsible}
\bibfield{author}{\bibinfo{person}{Bogdana Rakova}, \bibinfo{person}{Jingying
  Yang}, \bibinfo{person}{Henriette Cramer}, {and} \bibinfo{person}{Rumman
  Chowdhury}.} \bibinfo{year}{2020}\natexlab{}.
\newblock \showarticletitle{Where Responsible AI meets Reality: Practitioner
  Perspectives on Enablers for shifting Organizational Practices}.
\newblock \bibinfo{journal}{\emph{arXiv preprint arXiv:2006.12358}}
  (\bibinfo{year}{2020}).
\newblock


\bibitem[\protect\citeauthoryear{Rieman}{Rieman}{1993}]%
        {rieman1993diary}
\bibfield{author}{\bibinfo{person}{John Rieman}.}
  \bibinfo{year}{1993}\natexlab{}.
\newblock \showarticletitle{The diary study: a workplace-oriented research tool
  to guide laboratory efforts}. In \bibinfo{booktitle}{\emph{Proceedings of the
  INTERACT'93 and CHI'93 conference on Human factors in computing systems}}.
  \bibinfo{pages}{321--326}.
\newblock


\bibitem[\protect\citeauthoryear{Robertson, Deng, Gebru, Mitchell, Liebling,
  Lahav, Heller, D{\'\i}az, Bengio, and Salehi}{Robertson
  et~al\mbox{.}}{2021}]%
        {robertson_2021_directions}
\bibfield{author}{\bibinfo{person}{Samantha Robertson}, \bibinfo{person}{Wesley
  Deng}, \bibinfo{person}{Timnit Gebru}, \bibinfo{person}{Margaret Mitchell},
  \bibinfo{person}{Daniel~J Liebling}, \bibinfo{person}{Michal Lahav},
  \bibinfo{person}{Katherine Heller}, \bibinfo{person}{Mark D{\'\i}az},
  \bibinfo{person}{Samy Bengio}, {and} \bibinfo{person}{Niloufar Salehi}.}
  \bibinfo{year}{2021}\natexlab{}.
\newblock \showarticletitle{Three Directions for the Design of Human-Centered
  Machine Translation}.
\newblock \bibinfo{journal}{\emph{First Workshop on Bridging Human–Computer
  Interaction and Natural Language Processing at EACL 2021}}
  (\bibinfo{year}{2021}).
\newblock


\bibitem[\protect\citeauthoryear{Robertson and Díaz}{Robertson and
  Díaz}{2022}]%
        {robertson_2022_understanding}
\bibfield{author}{\bibinfo{person}{Samantha Robertson} {and}
  \bibinfo{person}{Mark Díaz}.} \bibinfo{year}{2022}\natexlab{}.
\newblock \showarticletitle{Understanding and Being Understood: User Strategies
  for Identifying and Recovering From Mistranslations in Machine
  Translation-Mediated Chat}. In \bibinfo{booktitle}{\emph{Proceedings of the
  2022 ACM Conference on Fairness, Accountability, and Transparency}} (Seoul,
  South Korea) \emph{(\bibinfo{series}{FAccT '22})}.
  \bibinfo{publisher}{Association for Computing Machinery},
  \bibinfo{address}{New York, NY, USA}.
\newblock


\bibitem[\protect\citeauthoryear{Sap, Gabriel, Qin, Jurafsky, Smith, and
  Choi}{Sap et~al\mbox{.}}{2019}]%
        {sap2019social}
\bibfield{author}{\bibinfo{person}{Maarten Sap}, \bibinfo{person}{Saadia
  Gabriel}, \bibinfo{person}{Lianhui Qin}, \bibinfo{person}{Dan Jurafsky},
  \bibinfo{person}{Noah~A Smith}, {and} \bibinfo{person}{Yejin Choi}.}
  \bibinfo{year}{2019}\natexlab{}.
\newblock \showarticletitle{Social bias frames: Reasoning about social and
  power implications of language}.
\newblock \bibinfo{journal}{\emph{arXiv preprint arXiv:1911.03891}}
  (\bibinfo{year}{2019}).
\newblock


\bibitem[\protect\citeauthoryear{Schilit, Adams, and Want}{Schilit
  et~al\mbox{.}}{1994}]%
        {schilit1994context}
\bibfield{author}{\bibinfo{person}{Bill Schilit}, \bibinfo{person}{Norman
  Adams}, {and} \bibinfo{person}{Roy Want}.} \bibinfo{year}{1994}\natexlab{}.
\newblock \showarticletitle{Context-aware computing applications}. In
  \bibinfo{booktitle}{\emph{1994 first workshop on mobile computing systems and
  applications}}. IEEE, \bibinfo{pages}{85--90}.
\newblock


\bibitem[\protect\citeauthoryear{Shigenobu}{Shigenobu}{2007}]%
        {shigenobu2007backtranslation}
\bibfield{author}{\bibinfo{person}{Tomohiro Shigenobu}.}
  \bibinfo{year}{2007}\natexlab{}.
\newblock \showarticletitle{Evaluation and Usability of Back Translation for
  Intercultural Communication}. In \bibinfo{booktitle}{\emph{Usability and
  Internationalization. Global and Local User Interfaces}},
  \bibfield{editor}{\bibinfo{person}{Nuray Aykin}} (Ed.).
  \bibinfo{publisher}{Springer Berlin Heidelberg}, \bibinfo{address}{Berlin,
  Heidelberg}, \bibinfo{pages}{259--265}.
\newblock
\showISBNx{978-3-540-73289-1}


\bibitem[\protect\citeauthoryear{Spechbach, Halimi~Mallem, Gerlach, Tsourakis,
  and Bouillon}{Spechbach et~al\mbox{.}}{2017a}]%
        {spechbach2017comparison}
\bibfield{author}{\bibinfo{person}{Hervé Spechbach},
  \bibinfo{person}{Ismahene~Sonia Halimi~Mallem}, \bibinfo{person}{Johanna
  Gerlach}, \bibinfo{person}{Nikolaos Tsourakis}, {and}
  \bibinfo{person}{Pierrette Bouillon}.} \bibinfo{year}{2017}\natexlab{a}.
\newblock \showarticletitle{Comparison of the quality of two speech translators
  in emergency settings : A case study with standardized Arabic speaking
  patients with abdominal pain}. In \bibinfo{booktitle}{\emph{Proceedings of
  European Congress of Emergency Medicine}} \emph{(\bibinfo{series}{EUSEM
  2017})}. \bibinfo{address}{Athens, Greece}.
\newblock
\urldef\tempurl%
\url{https://archive-ouverte.unige.ch/unige:100812}
\showURL{%
\tempurl}


\bibitem[\protect\citeauthoryear{Spechbach, Mallem, Gerlach, Tsourakis, and
  Bouillon}{Spechbach et~al\mbox{.}}{2017b}]%
        {Spechbach2017ComparisonOT}
\bibfield{author}{\bibinfo{person}{Herv{\'e} Spechbach},
  \bibinfo{person}{Ismahene Sonia~Halimi Mallem}, \bibinfo{person}{Johanna
  Gerlach}, \bibinfo{person}{Nikolaos Tsourakis}, {and}
  \bibinfo{person}{Pierrette Bouillon}.} \bibinfo{year}{2017}\natexlab{b}.
\newblock \showarticletitle{Comparison of the quality of two speech translators
  in emergency settings : A case study with standardized Arabic speaking
  patients with abdominal pain}.
\newblock


\bibitem[\protect\citeauthoryear{Specia, Shah, de~Souza, and Cohn}{Specia
  et~al\mbox{.}}{2013}]%
        {specia-etal-2013-quest}
\bibfield{author}{\bibinfo{person}{Lucia Specia}, \bibinfo{person}{Kashif
  Shah}, \bibinfo{person}{Jose~G.C. de Souza}, {and} \bibinfo{person}{Trevor
  Cohn}.} \bibinfo{year}{2013}\natexlab{}.
\newblock \showarticletitle{{Q}u{E}st - A translation quality estimation
  framework}. In \bibinfo{booktitle}{\emph{Proceedings of the 51st Annual
  Meeting of the Association for Computational Linguistics: System
  Demonstrations}}. \bibinfo{publisher}{Association for Computational
  Linguistics}, \bibinfo{address}{Sofia, Bulgaria}, \bibinfo{pages}{79--84}.
\newblock
\urldef\tempurl%
\url{https://www.aclweb.org/anthology/P13-4014}
\showURL{%
\tempurl}


\bibitem[\protect\citeauthoryear{Stecklow}{Stecklow}{2018}]%
        {stecklow2018facebook}
\bibfield{author}{\bibinfo{person}{Steve Stecklow}.}
  \bibinfo{year}{2018}\natexlab{}.
\newblock \showarticletitle{Why {Facebook} is losing the war on hate speech in
  {Myanmar}}.
\newblock \bibinfo{journal}{\emph{Reuters}} (\bibinfo{date}{Aug}
  \bibinfo{year}{2018}).
\newblock
\urldef\tempurl%
\url{https://www.reuters.com/investigates/special-report/myanmar-facebook-hate/}
\showURL{%
\tempurl}


\bibitem[\protect\citeauthoryear{Suresh, Gomez, Nam, and Satyanarayan}{Suresh
  et~al\mbox{.}}{2021}]%
        {suresh2021beyond}
\bibfield{author}{\bibinfo{person}{Harini Suresh}, \bibinfo{person}{Steven~R.
  Gomez}, \bibinfo{person}{Kevin~K. Nam}, {and} \bibinfo{person}{Arvind
  Satyanarayan}.} \bibinfo{year}{2021}\natexlab{}.
\newblock \showarticletitle{Beyond Expertise and Roles: A Framework to
  Characterize the Stakeholders of Interpretable Machine Learning and Their
  Needs}. In \bibinfo{booktitle}{\emph{Proceedings of the 2021 CHI Conference
  on Human Factors in Computing Systems}} (Yokohama, Japan)
  \emph{(\bibinfo{series}{CHI '21})}. \bibinfo{publisher}{Association for
  Computing Machinery}, \bibinfo{address}{New York, NY, USA}, Article
  \bibinfo{articleno}{74}, \bibinfo{numpages}{16}~pages.
\newblock
\showISBNx{9781450380966}
\urldef\tempurl%
\url{https://doi.org/10.1145/3411764.3445088}
\showDOI{\tempurl}


\bibitem[\protect\citeauthoryear{Thornton, Knowles, and Blair}{Thornton
  et~al\mbox{.}}{2021}]%
        {Thornton2021FiftySO}
\bibfield{author}{\bibinfo{person}{Lauren Thornton}, \bibinfo{person}{Bran
  Knowles}, {and} \bibinfo{person}{Gordon~S. Blair}.}
  \bibinfo{year}{2021}\natexlab{}.
\newblock \showarticletitle{Fifty Shades of Grey: In Praise of a Nuanced
  Approach Towards Trustworthy Design}.
\newblock \bibinfo{journal}{\emph{Proceedings of the 2021 ACM Conference on
  Fairness, Accountability, and Transparency}} (\bibinfo{year}{2021}).
\newblock


\bibitem[\protect\citeauthoryear{Torbati}{Torbati}{2019}]%
        {torbati2019immigration}
\bibfield{author}{\bibinfo{person}{Yeganeh Torbati}.}
  \bibinfo{year}{2019}\natexlab{}.
\newblock \showarticletitle{Google Says Google Translate Can’t Replace Human
  Translators. Immigration Officials Have Used It to Vet Refugees.}
\newblock \bibinfo{journal}{\emph{Pro Publica}} (\bibinfo{date}{September}
  \bibinfo{year}{2019}).
\newblock
\urldef\tempurl%
\url{https://www.propublica.org/article/google-says-google-translate-cant-replace-human-translators-immigration-officials-have-used-it-to-vet-refugees}
\showURL{%
\tempurl}


\bibitem[\protect\citeauthoryear{Translate}{Translate}{[n.d.]}]%
        {googletranslate}
\bibfield{author}{\bibinfo{person}{Google Translate}.}
  \bibinfo{year}{[n.d.]}\natexlab{}.
\newblock \bibinfo{booktitle}{\emph{Google Translation}}.
\newblock
\urldef\tempurl%
\url{https://translate.google.com/}
\showURL{%
\tempurl}


\bibitem[\protect\citeauthoryear{Turner, Choi, Dew, Tsai, Bosold, Wu, Smith,
  and Meischke}{Turner et~al\mbox{.}}{2019a}]%
        {turner2019evaluating}
\bibfield{author}{\bibinfo{person}{Anne~M Turner}, \bibinfo{person}{Yong~K
  Choi}, \bibinfo{person}{Kristin Dew}, \bibinfo{person}{Ming-Tse Tsai},
  \bibinfo{person}{Alyssa~L Bosold}, \bibinfo{person}{Shuyang Wu},
  \bibinfo{person}{Donahue Smith}, {and} \bibinfo{person}{Hendrika Meischke}.}
  \bibinfo{year}{2019}\natexlab{a}.
\newblock \showarticletitle{Evaluating the Usefulness of Translation
  Technologies for Emergency Response Communication: A Scenario-Based Study}.
\newblock \bibinfo{journal}{\emph{JMIR Public Health Surveill}}
  \bibinfo{volume}{5}, \bibinfo{number}{1} (\bibinfo{date}{Jan}
  \bibinfo{year}{2019}).
\newblock
\urldef\tempurl%
\url{https://doi.org/10.2196/11171}
\showDOI{\tempurl}


\bibitem[\protect\citeauthoryear{Turner, Choi, Dew, Tsai, Bosold, Wu, Smith,
  and Meischke}{Turner et~al\mbox{.}}{2019b}]%
        {Turner2019EvaluatingTU}
\bibfield{author}{\bibinfo{person}{Anne~M. Turner}, \bibinfo{person}{Yong~Kyung
  Choi}, \bibinfo{person}{Kristin~N. Dew}, \bibinfo{person}{Ming-Tse Tsai},
  \bibinfo{person}{Alyssa~L Bosold}, \bibinfo{person}{Shuyang Wu},
  \bibinfo{person}{Donahue Smith}, {and} \bibinfo{person}{Hendrika Meischke}.}
  \bibinfo{year}{2019}\natexlab{b}.
\newblock \showarticletitle{Evaluating the Usefulness of Translation
  Technologies for Emergency Response Communication: A Scenario-Based Study}.
\newblock \bibinfo{journal}{\emph{JMIR Public Health and Surveillance}}
  \bibinfo{volume}{5} (\bibinfo{year}{2019}).
\newblock


\bibitem[\protect\citeauthoryear{Vieira, O’Hagan, and O’Sullivan}{Vieira
  et~al\mbox{.}}{2020}]%
        {vieira2020societal}
\bibfield{author}{\bibinfo{person}{Lucas~Nunes Vieira}, \bibinfo{person}{Minako
  O’Hagan}, {and} \bibinfo{person}{Carol O’Sullivan}.}
  \bibinfo{year}{2020}\natexlab{}.
\newblock \showarticletitle{Understanding the societal impacts of machine
  translation: a critical review of the literature on medical and legal use
  cases}.
\newblock \bibinfo{journal}{\emph{Information, Communication \& Society}}
  \bibinfo{volume}{0}, \bibinfo{number}{0} (\bibinfo{year}{2020}),
  \bibinfo{pages}{1--18}.
\newblock
\urldef\tempurl%
\url{https://doi.org/10.1080/1369118X.2020.1776370}
\showDOI{\tempurl}
\showeprint{https://doi.org/10.1080/1369118X.2020.1776370}


\bibitem[\protect\citeauthoryear{Vieira, O’Hagan, and O’Sullivan}{Vieira
  et~al\mbox{.}}{2021}]%
        {Vieira21}
\bibfield{author}{\bibinfo{person}{Lucas~Nunes Vieira}, \bibinfo{person}{Minako
  O’Hagan}, {and} \bibinfo{person}{Carol O’Sullivan}.}
  \bibinfo{year}{2021}\natexlab{}.
\newblock \showarticletitle{Understanding the societal impacts of machine
  translation: a critical review of the literature on medical and legal use
  cases}.
\newblock \bibinfo{journal}{\emph{Information, Communication \& Society}}
  \bibinfo{volume}{24}, \bibinfo{number}{11} (\bibinfo{year}{2021}),
  \bibinfo{pages}{1515--1532}.
\newblock


\bibitem[\protect\citeauthoryear{William}{William}{2001}]%
        {William2001HarnessingPC}
\bibfield{author}{\bibinfo{person}{Colin~T. William}.}
  \bibinfo{year}{2001}\natexlab{}.
\newblock \showarticletitle{Harnessing People's Creativity: Ideation and
  Expression through Visual Communication}.
\newblock


\bibitem[\protect\citeauthoryear{Wu, Schuster, Chen, Le, Norouzi, Macherey,
  Krikun, Cao, Gao, Macherey, Klingner, Shah, Johnson, Liu, Kaiser, Gouws,
  Kato, Kudo, Kazawa, Stevens, Kurian, Patil, Wang, Young, Smith, Riesa,
  Rudnick, Vinyals, Corrado, Hughes, and Dean}{Wu et~al\mbox{.}}{2016}]%
        {Wu2016GooglesNM}
\bibfield{author}{\bibinfo{person}{Yonghui Wu}, \bibinfo{person}{Mike
  Schuster}, \bibinfo{person}{Z. Chen}, \bibinfo{person}{Quoc~V. Le},
  \bibinfo{person}{Mohammad Norouzi}, \bibinfo{person}{Wolfgang Macherey},
  \bibinfo{person}{Maxim Krikun}, \bibinfo{person}{Yuan Cao},
  \bibinfo{person}{Qin Gao}, \bibinfo{person}{Klaus Macherey},
  \bibinfo{person}{Jeff Klingner}, \bibinfo{person}{Apurva Shah},
  \bibinfo{person}{Melvin Johnson}, \bibinfo{person}{Xiaobing Liu},
  \bibinfo{person}{Lukasz Kaiser}, \bibinfo{person}{Stephan Gouws},
  \bibinfo{person}{Yoshikiyo Kato}, \bibinfo{person}{Taku Kudo},
  \bibinfo{person}{Hideto Kazawa}, \bibinfo{person}{Keith Stevens},
  \bibinfo{person}{George Kurian}, \bibinfo{person}{Nishant Patil},
  \bibinfo{person}{Wei Wang}, \bibinfo{person}{Cliff Young},
  \bibinfo{person}{Jason~R. Smith}, \bibinfo{person}{Jason Riesa},
  \bibinfo{person}{Alex Rudnick}, \bibinfo{person}{Oriol Vinyals},
  \bibinfo{person}{Gregory~S. Corrado}, \bibinfo{person}{Macduff Hughes}, {and}
  \bibinfo{person}{Jeffrey Dean}.} \bibinfo{year}{2016}\natexlab{}.
\newblock \showarticletitle{Google's Neural Machine Translation System:
  Bridging the Gap between Human and Machine Translation}.
\newblock \bibinfo{journal}{\emph{ArXiv}}  \bibinfo{volume}{abs/1609.08144}
  (\bibinfo{year}{2016}).
\newblock


\bibitem[\protect\citeauthoryear{Xu, Gao, Fussell, and Cosley}{Xu
  et~al\mbox{.}}{2014}]%
        {xu2014improving}
\bibfield{author}{\bibinfo{person}{Bin Xu}, \bibinfo{person}{Ge Gao},
  \bibinfo{person}{Susan~R. Fussell}, {and} \bibinfo{person}{Dan Cosley}.}
  \bibinfo{year}{2014}\natexlab{}.
\newblock \showarticletitle{Improving Machine Translation by Showing Two
  Outputs}. In \bibinfo{booktitle}{\emph{Proceedings of the SIGCHI Conference
  on Human Factors in Computing Systems}} (Toronto, Ontario, Canada)
  \emph{(\bibinfo{series}{CHI '14})}. \bibinfo{publisher}{Association for
  Computing Machinery}, \bibinfo{address}{New York, NY, USA},
  \bibinfo{pages}{3743–3746}.
\newblock
\showISBNx{9781450324731}
\urldef\tempurl%
\url{https://doi.org/10.1145/2556288.2557171}
\showDOI{\tempurl}


\bibitem[\protect\citeauthoryear{Yamashita, Inaba, Kuzuoka, and
  Ishida}{Yamashita et~al\mbox{.}}{2009a}]%
        {Yamashita2009DifficultiesIE}
\bibfield{author}{\bibinfo{person}{Naomi Yamashita}, \bibinfo{person}{Rieko
  Inaba}, \bibinfo{person}{Hideaki Kuzuoka}, {and} \bibinfo{person}{Toru
  Ishida}.} \bibinfo{year}{2009}\natexlab{a}.
\newblock \showarticletitle{Difficulties in establishing common ground in
  multiparty groups using machine translation}.
\newblock \bibinfo{journal}{\emph{Proceedings of the SIGCHI Conference on Human
  Factors in Computing Systems}} (\bibinfo{year}{2009}).
\newblock


\bibitem[\protect\citeauthoryear{Yamashita, Inaba, Kuzuoka, and
  Ishida}{Yamashita et~al\mbox{.}}{2009b}]%
        {yamashita2009difficulties}
\bibfield{author}{\bibinfo{person}{Naomi Yamashita}, \bibinfo{person}{Rieko
  Inaba}, \bibinfo{person}{Hideaki Kuzuoka}, {and} \bibinfo{person}{Toru
  Ishida}.} \bibinfo{year}{2009}\natexlab{b}.
\newblock \showarticletitle{Difficulties in Establishing Common Ground in
  Multiparty Groups Using Machine Translation}. In
  \bibinfo{booktitle}{\emph{Proceedings of the SIGCHI Conference on Human
  Factors in Computing Systems}} (Boston, MA, USA) \emph{(\bibinfo{series}{CHI
  '09})}. \bibinfo{publisher}{Association for Computing Machinery},
  \bibinfo{address}{New York, NY, USA}, \bibinfo{pages}{679–688}.
\newblock
\showISBNx{9781605582467}
\urldef\tempurl%
\url{https://doi.org/10.1145/1518701.1518807}
\showDOI{\tempurl}


\bibitem[\protect\citeauthoryear{Yamashita and Ishida}{Yamashita and
  Ishida}{2006}]%
        {Yamashita2006EffectsOM}
\bibfield{author}{\bibinfo{person}{Naomi Yamashita} {and} \bibinfo{person}{Toru
  Ishida}.} \bibinfo{year}{2006}\natexlab{}.
\newblock \showarticletitle{Effects of machine translation on collaborative
  work}. In \bibinfo{booktitle}{\emph{CSCW '06}}.
\newblock


\bibitem[\protect\citeauthoryear{Yang, Steinfeld, and Zimmerman}{Yang
  et~al\mbox{.}}{2019}]%
        {yang2019unremarkable}
\bibfield{author}{\bibinfo{person}{Qian Yang}, \bibinfo{person}{Aaron
  Steinfeld}, {and} \bibinfo{person}{John Zimmerman}.}
  \bibinfo{year}{2019}\natexlab{}.
\newblock \showarticletitle{Unremarkable ai: Fitting intelligent decision
  support into critical, clinical decision-making processes}. In
  \bibinfo{booktitle}{\emph{Proceedings of the 2019 CHI Conference on Human
  Factors in Computing Systems}}. \bibinfo{pages}{1--11}.
\newblock


\bibitem[\protect\citeauthoryear{Zimmerman and Forlizzi}{Zimmerman and
  Forlizzi}{2017}]%
        {zimmerman2017speed}
\bibfield{author}{\bibinfo{person}{John Zimmerman} {and} \bibinfo{person}{Jodi
  Forlizzi}.} \bibinfo{year}{2017}\natexlab{}.
\newblock \showarticletitle{Speed dating: providing a menu of possible
  futures}.
\newblock \bibinfo{journal}{\emph{She Ji: The Journal of Design, Economics, and
  Innovation}} \bibinfo{volume}{3}, \bibinfo{number}{1} (\bibinfo{year}{2017}),
  \bibinfo{pages}{30--50}.
\newblock


\bibitem[\protect\citeauthoryear{Zouhar, Nov{\'a}k, {\v{Z}}ilinec, Bojar,
  Obreg{\'o}n, Hill, Blain, Fomicheva, Specia, and Yankovskaya}{Zouhar
  et~al\mbox{.}}{2021}]%
        {zouhar-etal-2021-backtranslation}
\bibfield{author}{\bibinfo{person}{Vil{\'e}m Zouhar}, \bibinfo{person}{Michal
  Nov{\'a}k}, \bibinfo{person}{Mat{\'u}{\v{s}} {\v{Z}}ilinec},
  \bibinfo{person}{Ond{\v{r}}ej Bojar}, \bibinfo{person}{Mateo Obreg{\'o}n},
  \bibinfo{person}{Robin~L. Hill}, \bibinfo{person}{Fr{\'e}d{\'e}ric Blain},
  \bibinfo{person}{Marina Fomicheva}, \bibinfo{person}{Lucia Specia}, {and}
  \bibinfo{person}{Lisa Yankovskaya}.} \bibinfo{year}{2021}\natexlab{}.
\newblock \showarticletitle{Backtranslation Feedback Improves User Confidence
  in {MT}, Not Quality}. In \bibinfo{booktitle}{\emph{Proceedings of the 2021
  Conference of the North American Chapter of the Association for Computational
  Linguistics: Human Language Technologies}}. \bibinfo{publisher}{Association
  for Computational Linguistics}, \bibinfo{address}{Online},
  \bibinfo{pages}{151--161}.
\newblock
\urldef\tempurl%
\url{https://doi.org/10.18653/v1/2021.naacl-main.14}
\showDOI{\tempurl}


\end{thebibliography}
